\documentclass[journal]{IEEEtran}

\usepackage{amsmath,amsfonts}
\usepackage{algorithmic}
\usepackage{algorithm}
\usepackage{array}
\usepackage[caption=false,font=normalsize,labelfont=sf,textfont=sf]{subfig}
\usepackage{textcomp}
\usepackage{stfloats}
\usepackage{url}
\usepackage{verbatim}
\usepackage{graphicx}
\usepackage{cite}

\usepackage[hidelinks]{hyperref}
\usepackage[table,xcdraw]{xcolor}
\usepackage{pifont} %

\usepackage{booktabs}
\usepackage{multirow}
\usepackage{makecell}

\usepackage{FiraMono}
\usepackage[T1]{fontenc}

\usepackage{listings}
\definecolor{codeBackground}{RGB}{255,255,255}
\definecolor{codeComment}{RGB}{0,156,0}
\definecolor{codeKeyword}{RGB}{0,0,255}
\definecolor{codeLiteral}{RGB}{233,0,0}
\definecolor{codeString}{RGB}{233,0,0}
\definecolor{codeDefault}{RGB}{0,0,0}
\definecolor{codeNumbers}{RGB}{150,150,150}
\lstdefinelanguage{HTML5}{
  language=html,
  sensitive=true,
  alsodigit={.:;},
  alsoletter={<>=-},
  morecomment=[s]{<!-}{-->},
  tag=[s],
  escapeinside={<@}{@>},
  otherkeywords={
  >,
  <!DOCTYPE, <!doctype,
  </html, <html, <head, <title, </title, <style, </style, <link, </head, <meta, />,
  </body, <body,
  </div, <div, </div>,
  </p, <p, </p>, </h1, <h1,
  </script, <script,
  <canvas, /canvas>, <svg, <rect, <animateTransform, </rect>, </svg>, <video, <source, <iframe, </iframe>, </video>, <image, </image>, <header, </header, <article, </article
  },
  ndkeywords={
  =,
  charset=, src=, id=, width=, height=, style=, type=, rel=, href=,
  fill=, attributeName=, begin=, dur=, from=, to=, poster=, controls=, x=, y=, repeatCount=, xlink:href=,
  margin:, padding:, background-image:, border:, top:, left:, position:, width:, height:, margin-top:, margin-bottom:, font-size:, font-family:, line-height:,
  transform:, -moz-transform:, -webkit-transform:,
  animation:, -webkit-animation:,
  transition:,  transition-duration:, transition-property:, transition-timing-function:,
  }
}
\lstdefinelanguage{JavaScript}{
  keywords={typeof, new, true, false, catch, function, return, null, catch, switch, var, if, in, while, do, else, case, break},
  ndkeywords={class, export, boolean, throw, implements, import, this},
  keywordstyle=\color{codeKeyword}\bfseries,
  ndkeywordstyle=\color{codeKeyword}\bfseries,
  commentstyle=\color{codeComment}\ttfamily,
  stringstyle=\color{codeString}\ttfamily,
  identifierstyle=\color{codeDefault},
  sensitive=false,
  comment=[l]{//},
  morecomment=[s]{/*}{*/},
  morestring=[b]',
  morestring=[b]"
}
\usepackage{xspace}
\usepackage{relsize}
\newcommand\eg{e.g.,\xspace}
\newcommand\ie{i.e.,\xspace}
\newcommand\etal{et al.\xspace}
\newcommand{\code}[1]{\smash{\smaller[0.5]\lstinline[language=JavaScript]{#1}}}

\usepackage{framed}

\definecolor{blue}{HTML}{0c84ed}
\definecolor{orange}{HTML}{ff8811}
\definecolor{purple}{HTML}{3b0086}
\newcommand\datasetExtensionsTotal{1,000\xspace}
\newcommand\datasetExtensionsOk{941\xspace}
\newcommand\datasetExtensionEntrypointsOk{6,056\xspace}
\newcommand\datasetCdnjsTotal{4,135\xspace}
\newcommand\datasetCdnjsOk{4,006\xspace}

\newcommand\datasetExtensionEntrypointsDynamicCrashP{38\%\xspace}

\newcommand\datasetExtensionEntrypointsDynamicTimeoutP{4\%\xspace}

\newcommand\datasetCdnjsDynamicCrashP{17\%\xspace}

\newcommand\datasetCdnjsDynamicTimeoutP{1\%\xspace}

\newcommand\perfStaticTimeAvg{1,201\xspace}
\newcommand\perfStaticTimeStd{10,564\xspace}

\newcommand\perfStaticMemoryAvg{67.5\xspace}
\newcommand\perfStaticMemoryStd{273.3\xspace}
\newcommand\perfStaticMemoryMax{9902.2\xspace}
\newcommand\perfDynamicTimeAvg{70\xspace}
\newcommand\perfDynamicTimeStd{373\xspace}

\newcommand\perfDynamicTimeQNN{728\xspace}
\newcommand\perfDynamicMemoryAvg{1.5\xspace}
\newcommand\perfDynamicMemoryStd{1.8\xspace}
\newcommand\perfDynamicMemoryMin{0.9\xspace}
\newcommand\perfDynamicMemoryMax{120.7\xspace}

\newcommand\apiCallsStaticAvgStd{43$\pm$67\xspace}
\newcommand\apiCallsDynamicAvgStd{8$\pm$14\xspace}
\newcommand\apiCallsHiddenAvgStd{4$\pm$10\xspace}
\newcommand\apiCallsEmptyExtensions{205\xspace}
\newcommand\apiCallsEmptyLibraries{82\xspace}
\newcommand\apiCallsNotEmptySources{9,032\xspace}

\newcommand\apiCallsSourcesWithHiddenP{57\%\xspace}

\newcommand\stringsStaticAvg{1,221\xspace}
\newcommand\stringsDynamicAvg{91\xspace}

\newcommand\stringsSourcesWithHiddenP{41\%\xspace}

\newcommand\stringHiddenTotal{205,919\xspace}
\newcommand\stringHiddenIpvFour{18\xspace}
\newcommand\stringHiddenUuid{32\xspace}
\newcommand\stringHiddenUrls{339\xspace}
\newcommand\stringHiddenUrlsOrigins{175\xspace}
\newcommand\stringHiddenUrlsGa{84\xspace}
\newcommand\stringHiddenJson{422\xspace}
\newcommand\stringHiddenJsonUuid{22\xspace}

\begin{document}

\title{Fakeium: A Dynamic Execution Environment for JavaScript Program Analysis}

\author{
  José Miguel Moreno, %
  \thanks{
    José Miguel Moreno and Juan Tapiador are with the Universidad Carlos III de Madrid, 28911 Madrid, Spain
    (e-mail: josemore@pa.uc3m.es, jestevez@inf.uc3m.es).
    Narseo Vallina-Rodriguez is with the IMDEA Networks Institute, 28918 Madrid, Spain
    (e-mail: narseo.vallina@imdea.org).
  }%
  Narseo Vallina-Rodriguez, and %
  Juan Tapiador,~\IEEEmembership{Senior Member, IEEE}
  \thanks{
    This research was supported by the Spanish AEI grant CYCAD (PID2022-140126OB-I00).
    José Miguel Moreno is funded by the Spanish Ministry of Science and Innovation with a FPI Predoctoral Grant (PRE2020-094224).
    Narseo Vallina-Rodriguez has been appointed as 2019 Ramon y Cajal fellow (RYC2020-030316-I) funded by MICIU/AEI/10.13039/501100011033 and the ESF Investing in your future.
  }
}

\maketitle

\begin{abstract}
The JavaScript programming language, which began as a simple scripting language for the Web, has become ubiquitous, spanning desktop, mobile, and server applications.
This increase in usage has made JavaScript an attractive target for nefarious actors, resulting in the proliferation of malicious browser extensions that steal user information and supply chain attacks that target the official Node.js package registry.
To combat these threats, researchers have developed specialized tools and frameworks for analyzing the behavior of JavaScript programs to detect malicious patterns.
Static analysis tools typically struggle with the highly dynamic nature of the language and fail to process obfuscated sources, while dynamic analysis pipelines take several minutes to run and require more resources per program, making them unfeasible for large-scale analyses.
In this paper, we present Fakeium, a novel, open source, and lightweight execution environment designed for efficient, large-scale dynamic analysis of JavaScript programs.
Built on top of the popular V8 engine, Fakeium complements traditional static analysis by providing additional API calls and string literals that would otherwise go unnoticed without the need for resource-intensive instrumented browsers or synthetic user input.
Besides its negligible execution overhead, our tool is highly customizable and supports hooks for advanced analysis scenarios such as network traffic emulation.
Fakeium's flexibility and ability to detect hidden API calls, especially in obfuscated sources, highlights its potential as a valuable tool for security analysts to detect malicious behavior.
\end{abstract}


\section{Introduction}
\label{sec:introduction}

\IEEEPARstart{S}{ince} its creation in late 1995, JavaScript has grown to become one of the most ubiquitous programming languages, shaping modern software development and even inspiring other alternatives like CoffeeScript and, more recently, TypeScript.
While originally designed to add interactivity to early web browsers, JavaScript now powers a wide range of applications beyond webpages.
These include browser extensions that enhance or customize the behavior and UI of the browser; backend services and serverless functions that run under Node.js or Deno; desktop applications powered by Electron, WebView2, or similar runtimes; and cross-platform mobile apps built with React Native or leveraging Android's WebView and Apple's WKWebView.

As its use has expanded to cover almost all platforms, so does the potential for misuse.
Malicious browser extensions have been found harvesting personal user information on numerous occasions, in some cases even being distributed through official channels like the Chrome Web Store~\cite{report-palant-malicious-extensions,report-kaspersky-dangerous-extensions,report-cashback-extension-killer,report-extensions-dll-malware}.
Popular Node.js packages found in the npm registry have also been targeted by malicious actors in supply chain attacks to distribute cryptominers or steal sensitive credentials~\cite{report-npm-north-korea,report-npm-cryptomining,report-npm-noblox}.
These affected packages are then included as dependencies in server, desktop, and mobile applications, further extending the impact of these kinds of campaigns.
With such widespread adoption and the increasing frequency of these attacks, the development of modern and effective tools for analyzing JavaScript programs is critical to the research community.
This is important not only for security and threat intelligence professionals to find malware and vulnerable programs, but also in the privacy field to detect information leaks and fingerprinting attempts.

The automated analysis of JavaScript code has been explored before, both using static and dynamic approaches~\cite{tajs-first,actarus,nodest,mininode,jstap,hulk,fv8,arcanum}.
However, developing and maintaining such tools is challenging due to the highly dynamic nature of the JavaScript language~\cite{survey-of-dynamic-analysis}.
For starters, static analysis struggles to extract signals (such as API calls and strings) that are only present at runtime.
This is particularly important in the security context, where obfuscated code that hides the malicious payload of a program is common.
When used for privacy analysis, static techniques may also miss fingerprinting methods designed to evade detection.
Dynamic analysis is often used to overcome these limitations, and involves running the program in an instrumented environment to log meaningful signals that help analysts understand its behavior.
Unfortunately, traditional dynamic analysis also comes with its own set of challenges, as it requires user input (manual or automated) to trigger a particular code path within the analyzed program.
Most importantly, it has a critical disadvantage when compared to static analysis: it is computationally expensive and time consuming, making it unfeasible for large-scale analyses of millions of sources.

To address these challenges, we present Fakeium, a novel, open source, and lightweight dynamic execution environment specifically designed for inferring the behavior of JavaScript programs at scale.
Unlike previous works, Fakeium is highly customizable through built-in support for hooks, and its dynamic nature makes it ideal for analyzing obfuscated sources.
By focusing on efficiency and scalability, our tool offers analysts an automated method of analyzing sources without the need for synthetic user input or resource-intensive Chromium browser instances.
Fakeium does not intend to replace static analysis.
Instead, it aims to complement it by providing further and more reliable signals in cases where traditional dynamic analysis is not possible due to resource and time constrains.
All this while having a minimal overhead over static analysis of mere \perfDynamicTimeAvg ms and \perfDynamicMemoryAvg MiB of heap on average.
We evaluate our tool on a diverse dataset of browser extensions and JavaScript libraries and find it capable of detecting API calls missed by static analysis in \apiCallsSourcesWithHiddenP of sources.
We also showcase Fakeium's customization capabilities by analyzing a recent malicious sample, revealing its behavior despite its heavy obfuscation.

\vspace{1mm}
\noindent\textbf{Software Release.}\xspace
We distribute Fakeium as a Node.js package that can be easily installed using the \code{npm install fakeium} command.
We also release its source code under the MIT license at \url{https://github.com/josemmo/fakeium} in the hope that it will contribute to the advancement of research tools for security analysis of JavaScript programs.

\section{Background}
\label{sec:background}

The analysis of JavaScript applications is still an open problem that has been worked on for more than a decade.
As with other programming languages and platforms, existing techniques have been broadly divided into two major groups: static and dynamic analysis, each one with its own application domain and limitations.
In this section, we outline the challenges and limitations of current state-of-the-art analysis techniques.
We also discuss relevant considerations and programming language concepts which are necessary for an adequate understanding of these analysis methods.

\subsection{Language Considerations}
JavaScript is a high-level programming language widely used in web development, both client and server side, as well as desktop and mobile applications.

It is based on the ECMAScript Language Specification~\cite{ecmascript}.
A popular engine that implements this specification is V8, which is used by all Chromium-based browsers and the Node.js runtime~\cite{v8}.
For code maintainability, JavaScript programs can be split into reusable and isolated \textit{modules}.
When a part of the program needs a particular function or variable from a different module, it can be imported using the \code{import} language keyword.
This variable isolation is provided through \textit{scopes}.
In JavaScript, a scope is an execution context that isolates which variables are accessible at a particular part of the program execution~\cite{mdn-scope}.
A given piece of code cannot interact with variables outside its scope, except for the global scope (\code{globalThis}), which is always accessible from anywhere in the program.

\subsection{Static Analysis}
\label{sec:background:static-analysis}
Performing static analysis on a JavaScript program involves examining its source code without executing it.

Parsing the source code directly is challenging due to the complexity of ECMAScript syntax and the highly dynamic nature of the language~\cite{survey-of-dynamic-analysis}.
Instead, static analysis tools most commonly work with Abstract Syntax Trees (ASTs), a higher-level abstraction of the program generated from the source code.
ASTs are intermediate representations of a program whose nodes can be traversed to extract information or safely apply transformations.
AST nodes may be enhanced with contextual information extracted by the parsing tool, such as the number of times a variable is referenced, which is useful for performing optimizations on the program like dead code removal or \textit{tree shaking}~\cite{mdn-tree-shaking}.
Popular tools that allow developers to build and work with JavaScript ASTs are Acorn~\cite{acorn}, Babel~\cite{babeljs} and Esprima~\cite{esprima}.

For other platforms, such as Android, static analyzers typically do not use ASTs, relying instead on other structures like Control-Flow Graphs (CFGs)~\cite{appangio, eight-years-android, glaciate}.

An important limitation of these tools is that they can only produce ASTs of isolated sources.
That is, they do not implement module resolution.
This reduces the detection of API calls as shown in Listing~\ref{lst:ast-limitations}, where there is not enough context to log the call to \code{navigator.geolocation.getCurrentPosition} when looking at the two trees separately.
To get around it, researchers can combine multiple JavaScript sources in a process called \textit{bundling} and then obtain the AST of the resulting bundle~\cite{did-i-vet-you-before}.
Bundling is widely used in the web ecosystem to merge codebases of thousands of source files and assets into a single file (or a few files) ready for production distribution.
Some popular bundling tools include esbuild~\cite{esbuild}, Rollup~\cite{rollup} and Webpack~\cite{webpack}.

\begin{lstlisting}[
  float=t,
  caption={Example code to illustrate the limitations of AST generation without module resolution.},
  label={lst:ast-limitations},
  language=JavaScript
]
// deps.mjs
export function getGeolocation() {
  return navigator.geolocation;
}

// index.mjs
import { getGeolocation } from "./deps.mjs";
getGeolocation().getCurrentPosition(/* [...] */);
\end{lstlisting}

Another and more difficult challenge with static analysis comes with dynamically executed and highly obfuscated code.
JavaScript lets developers evaluate code coming from a string through the use of the \code{eval} function and the \code{Function} constructor~\cite{exploring-js-dynamic-execution}.
Babel and other parsers translate this behavior in the AST as a \code{CallExpression} with a \code{Literal} node argument containing the source code string, without recursively parsing its contents by default.
Thus, malicious actors may use these techniques to evade detection.
Furthermore, both free and commercial JavaScript code obfuscators are easily accessible~\cite{javascript-obfuscator, jsdefender}.
Deobfuscating programs produced by these tools requires the development of specific utilities for each obfuscator, which involves considerable effort and maintenance.
Obfuscation is such a major concern that the Chrome Web Store has banned browser extensions that have obfuscated code~\cite{cws-policies-readability}.

All in all, static analysis has multiple advantages.
For starters, it is generally fast, with analyses completing in the order of seconds.
In addition, existing tools have proven to be reliable and widely used in the web industry, most notably Babel, which is funded by large companies such as Airbnb, Salesforce, and Facebook~\cite{babeljs}.
Lastly, static analysis is suitable for determining all \textit{possible} behaviors of a program, even if they are not exhibited at runtime.
Note, however, that the use of obfuscation and techniques for dynamically evaluating code hampers the ability of static analysis to identify possible behaviors.

For example, a website may choose to display a consent cookie banner to visitors from the EU only.
This trait is valuable for finding hidden code paths and extracting signals that can later be fed into a dynamic analysis pipeline.

\subsection{Dynamic Analysis}
In contrast to static analysis, dynamic analysis aims to extract information about a program by running it.
This is accomplished by executing the program in an instrumented environment or \textit{sandbox}, where calls to certain APIs, memory allocations, network traffic, and other resources are monitored.
While there are several general purpose sandboxes for malware analysis~\cite{joe-sandbox, hybrid-analysis}, there is no dynamic execution environment specifically designed for running JavaScript code.
Instead, researchers typically create their own sandboxes by instrumenting a Chromium-based browser instance with end-to-end testing tools like Playwright, Puppeteer or Selenium~\cite{playwright, puppeteer, selenium}.
Although these tools are not intended for malware analysis, their flexibility and powerful capabilities make them good candidates for the task.

In almost all cases, dynamic analysis requires some kind of user input to trigger or \textit{elicit} a certain behavior in the program.

For example, some browser extensions may refuse to run until the user agrees to their privacy policy, or may only work on certain websites.
One solution to this problem is to have a human interact with the sandbox, providing accurate input that elicits as many behaviors in the program as possible.
Unfortunately, introducing a human-in-the-loop does not scale to thousands or hundreds of analyses.
Instead, another common approach is to fully automate the analysis by using a tool that provides synthetic input.
These exerciser tools are informally called \textit{monkeys}, a borrowed terminology from Android development~\cite{android-monkey}.
Eliciting behavior is not a trivial task, often requiring high-quality signals (such as URLs and API calls) previously extracted from static analysis.

Compared to static analysis, a major advantage of dynamic analysis is that it can analyze obfuscated and dynamically executed code.
However, this comes at the expense of more computing resources to run the sandbox, and time to feed the user input and wait for feedback.
In addition, dynamic analysis does not necessarily cover all possible behaviors of a program, but only those expressed in a particular execution.
This means that certain code paths hidden behind signals (such as a URL the user must visit), time bombs, or other protections may be missed by the sandbox.

\section{Related Work}
\label{sec:related-work}

The development of effective tools for the analysis of JavaScript programs has been a subject of focus for the web security community, as shown by the surveys from Sun and Ryu~\cite{survey-of-static-analysis}, and Andreasen \etal~\cite{survey-of-dynamic-analysis}.

\vspace{1mm}
\noindent\textbf{Static Analysis Tools.}\xspace
Jensen \etal developed TAJS, a type analyzer for JavaScript that focuses on helping developers catch programming bugs~\cite{tajs-first, tajs-second, tajs-third}.
Guarnieri \etal developed \textsc{Actarus}, a tool with a built-in taint analysis algorithm for finding vulnerabilities in client-side JavaScript~\cite{actarus}.
Kashyap \etal defined an intermediate language to create JSAI, an abstract interpreter for JavaScript code that aims to be more accurate than the existing TAJS~\cite{jsai}.
Nielsen \etal introduced \textsc{Nodest}, a tool built on top of TAJS for finding vulnerable taint flows in server-side applications~\cite{nodest}.
Similarly, Koishybayev and Kapravelos developed Mininode, a tool for detecting and removing unused dependencies to reduce the attack surface of Node.js applications~\cite{mininode}.
Fass \etal introduced \textsc{JStap}, a modular tool that uses machine learning to classify between malicious and benign JavaScript samples~\cite{jstap}.

\vspace{1mm}
\noindent\textbf{Dynamic Analysis Tools.}\xspace
Kapravelos \etal introduced Hulk, a dynamic analysis system that relies on fuzzing to elicit malicious behaviors in browser extensions~\cite{hulk}.
Taking a different approach, Kim \etal used forced execution to develop \textsc{J-Force}, a JavaScript engine implemented using WebKit to find malicious behaviors in webpages or extensions~\cite{jforce}.
Instead of creating an entire new engine, Li \etal modified the Chromium browser to extend the instrumentation capabilities provided by the Chrome DevTools, naming their tool JSgraph~\cite{jsgraph}.
In the same fashion, Jueckstock and Kapravelos created VisibleV8, a modified version of the V8 engine that logs accesses to native functions and properties to detect evasive code~\cite{visiblev8}.
More recently, Pantelaios and Kapravelos developed another custom V8 build with forced execution named FV8~\cite{fv8}; and Kie \etal created Arcanum, a fork of Chromium that implements dynamic taint tracking for browser extensions~\cite{arcanum}.

\vspace{1mm}
\noindent\textbf{Obfuscation.}\xspace
Sarker \etal used VisibleV8 and a mix of static and dynamic analysis to detect obfuscated JavaScript sources in the top 100k domains from Alexa~\cite{hiding-in-plain-sight}.
Also looking at the Alexa ranking, Moog \etal used static analysis with Esprima and \textsc{JStap} to detect both minified and obfuscated code~\cite{js-obfs-moog}.
Ren \etal developed JSRevealer, a tool for detecting malicious behavior that is resistant to obfuscation~\cite{jsreleaver}.

Table~\ref{tab:tools-comparison} shows a comparison between the aforementioned proposals and ours.
Of all the tools studied, Fakeium is the only one that is under active development (\ie has received an update in the last year) and makes an effort to ensure the quality of its code by both linting and testing it.
As opposed to previous work, our tool provides support for extensive customization such as hooks and sandbox execution limits without the need to modify its source code.
Furthermore, Fakeium is easy to install as it is distributed as an npm package, and is well documented.

\begin{table}[t]
  \centering
  \caption{Comparison of related work with our tool.}
  \label{tab:tools-comparison}
  \newcommand\rot[1]{\rotatebox{90}{#1}}
  \newcommand\good[1]{\textcolor{green!70!black}{#1}}
  \newcommand\meh[1]{\textcolor{orange!90!black}{#1}}
  \newcommand\bad[1]{\textcolor{red!80!black}{#1}}
  \newcommand\yes{\good{\ding{51}}}
  \newcommand\no{\bad{\ding{55}}}
  \newcommand\unk{\meh{\textbf{--}}}
    \begin{tabular}{lllcccc}
      \toprule
      \thead[l]{Tool}                 & \thead[l]{Targets} & \thead[l]{License} & \thead[c]{\rot{Active}} & \thead[c]{\rot{Linted}} & \thead[c]{\rot{Tested}} & \thead[c]{\rot{Cust.}} \\
      \midrule
      TAJS~\cite{tajs-first}          & \good{Any}         & \good{Apache}      & \no                     & \no                     & \yes                    & \no                    \\
      \textsc{Actarus}~\cite{actarus} & \meh{Web}          & \bad{Proprietary}  & \yes                    & \unk                    & \unk                    & \unk                   \\
      JSAI~\cite{jsai}                & \good{Any}         & \meh{No license}   & \no                     & \no                     & \no                     & \no                    \\
      \textsc{Nodest}~\cite{nodest}   & \meh{Node.js}      & \bad{Proprietary}  & \unk                    & \unk                    & \unk                    & \unk                   \\
      Mininode~\cite{mininode}        & \meh{Node.js}      & \good{BSD}         & \no                     & \no                     & \yes                    & \no                    \\
      \textsc{JStap}~\cite{jstap}     & \good{Any}         & \good{AGPLv3}      & \no                     & \no                     & \no                     & \no                    \\
      Hulk~\cite{hulk}                & \meh{Extensions}   & \bad{Proprietary}  & \unk                    & \unk                    & \unk                    & \unk                   \\
      \textsc{J-Force}~\cite{jforce}  & \good{Any}         & \bad{Proprietary}  & \unk                    & \unk                    & \unk                    & \unk                   \\
      JSgraph~\cite{jsgraph}          & \meh{Extensions}   & \good{GPLv2}       & \no                     & \no                     & \no                     & \no                    \\
      VisibleV8~\cite{visiblev8}      & \meh{Web}          & \good{BSD}         & \yes                    & \no                     & \yes                    & \no                    \\
      FV8~\cite{fv8}                  & \good{Any}         & \good{MIT}         & \yes                    & \no                     & \yes                    & \no                    \\
      Arcanum~\cite{arcanum}          & \meh{Extensions}   & \good{MIT}         & \yes                    & \no                     & \no                     & \no                    \\
      \midrule
      Fakeium                         & \good{Any}         & \good{MIT}         & \yes                    & \yes                    & \yes                    & \yes                   \\
      \bottomrule
    \end{tabular}
\end{table}

\section{Fakeium}
\label{sec:fakeium}

Fakeium (short for \textit{Fake Chromium}) is a lightweight, general-purpose sandbox for the dynamic execution of untrusted JavaScript code.

In this section, we outline the design goals for this project and describe the architecture and implementation of the Fakeium sandbox.

\subsection{Design Goals}
Given the time and computational constraints of classical JavaScript sandboxes, it is not always possible to perform dynamic analysis at scale.
Thus, our main goal is to fill this technological gap by creating a tool that overcomes the limitations of static analysis when traditional dynamic analysis is not feasible.

We note that we are not trying to replace static analysis, but to complement it with better signals that are difficult or impossible to find with static analysis alone.

Other secondary goals include:

\vspace{1mm}
\noindent\textbf{Efficiency.}\xspace
We aim for Fakeium to have a minimal impact in terms of execution time, so that it can be run alongside a static analyzer without significant overhead.
We also target low CPU usage and a small memory footprint by keeping the sandbox as lightweight as possible.

\vspace{1mm}
\noindent\textbf{Usability.}\xspace
Fakeium is expected to work out of the box, without requiring any monkey, user input or signals extracted from static analysis.
Also for usability, we decided to distribute it as an npm library written in TypeScript, so that developers get improved type hinting even when programming in JavaScript.

\vspace{1mm}
\noindent\textbf{Extensibility.}\xspace
To account for execution environments that need a higher level of customization (such as proxying real network traffic on calls to \code{fetch} and \code{XMLHttpRequest}), we designed Fakeium with support for custom hooks in mind.

\vspace{1mm}
\noindent\textbf{Correctness.}\xspace
In view of the complexity of the JavaScript language, we find it necessary to write an extensive suite of unit tests to ensure accurate execution output, as well as to minimize potential regression bugs that may appear during extended development.

\subsection{Architecture}
\label{sec:fakeium:architecture}

\begin{figure}[t]
  \centering
  \includegraphics[width=\columnwidth]{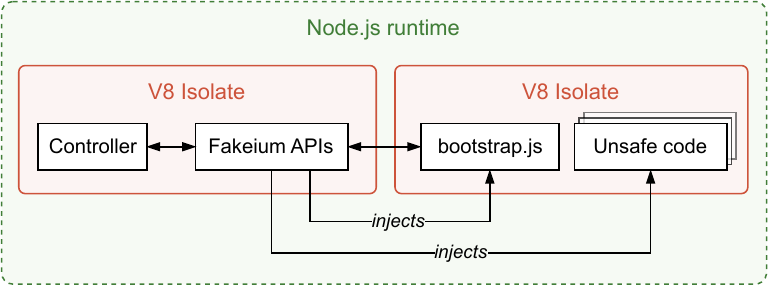}
  \caption{Overview of Fakeium architecture.}
  \label{fig:architecture}
\end{figure}

\begin{lstlisting}[
  float=t,
  caption={Simple example program for executing code in the Fakeium sandbox.},
  label={lst:fakeium-example},
  language=JavaScript
]
import { Fakeium } from "fakeium";

(async () => {
  // Setup instance
  const fakeium = new Fakeium({
    sourceType: "script",
    origin: "https://example.com/",
  });
  await fakeium.run("specifier.js", "alert(1)");

  // Print all logged events
  const events = fakeium.getReport().getAll();
  console.dir(events, { depth: null });

  // Free resources
  fakeium.dispose();
})();
\end{lstlisting}

Figure~\ref{fig:architecture} shows an overview of the architecture used by Fakeium.
Developers write a Node.js program (\ie controller) to interact with the Fakeium APIs.
These publicly exposed APIs enable them to instrument the sandbox, run code in it, and get reports on its execution (see Listing~\ref{lst:fakeium-example} for an example).
To safely run untrusted code, Fakeium uses the isolated-vm library~\cite{isolated-vm}.
This library allows us to hook into the V8 engine internals to create separate \code{v8::Isolate}s that run under the same Node.js process.
An \textit{isolate} is an instance of the V8 engine that has its own heap and associated objects~\cite{v8-embedded-builtins}.
All code must belong to an isolate and, by default, Node.js creates a single one shared throughout the execution of a program.
Since changes made to an isolate do not affect others, we can safely spawn an isolate, run untrusted code in it, and then dispose it.
Furthermore, we can limit the amount of memory it can allocate and terminate its execution after an overrun or timeout.

The actual Fakeium sandbox (where the untrusted code runs) is a separate V8 isolate with a barebones execution context.
That is, the sandbox natively exposes only the built-in objects and functions provided by V8 according to the ECMAScript specification (\eg \code{Error}, \code{JSON}, \code{eval}).
Neither the Document Object Model (DOM) nor APIs that would allow the potentially harmful code to interact with external resources (like \code{fetch} or Node.js APIs) are exposed for safety reasons.

Before running any user-provided code in the sandbox, Fakeium prepares the context of the newly created isolate by executing a bootstrapping script.
The role of \textit{bootstrap.js} is to hijack the global scope inside the isolate and report the occurring events back to the Fakeium APIs outside the sandbox.
This communication is achieved by passing references to callbacks from the external isolate as arguments to the bootstrapping script.
Because this script is evaluated as a closure, the unsafe code that runs in the isolate after it cannot access the callback references.
To account for the missing APIs that the code being analyzed might expect, we mock all accessed variables at runtime.
We explain mocks in more detail in Section~\ref{sec:fakeium:implementation}.

\subsection{Implementation Details}
\label{sec:fakeium:implementation}

Fakeium is implemented in 1.8k lines of code, excluding comments and whitespace.
We use ESLint with its strict and stylistic configurations for linting~\cite{eslint}, and an extensive suite of 40+ unit tests to ensure runtime correctness.
We use pkgroll to build the project~\cite{pkgroll}, resulting in a final unpacked size of only 50 kiB.

\vspace{1mm}
\noindent\textbf{Mocks.}\xspace
One of the key features of Fakeium is that it \textit{mocks} or creates dummies for objects that are missing from the global scope at runtime.
Mocks are created recursively as soon as the analyzed program requests an undefined property (\eg \code{chrome.cookies.getAll}).
The main advantage of this approach is that we do not need to implement all of the existing JavaScript APIs, nor we need to know in advance what variables the program expects.
Another advantage is that mocks are deterministic: since they do not rely on external resources, the report of a program analyzed with Fakeium will be the same after several runs.\footnote{
  The exception to this rule is the ECMAScript specification of \code{Math.random}.
  Note, however, that Fakeium enables developers to change this behavior using custom hooks, as we explain in the next subsection.
}
The trade-off with mocks is that they provide execution times orders of magnitude faster than traditional dynamic analysis sandboxes at the expense of accuracy.
We still see this is as a worthwhile compromise, as Fakeium seeks to be a good enough replacement for dynamic analysis when the latter is not feasible.

Mocks are implemented using \code{Proxy} objects~\cite{mdn-proxy} by defining traps for all relevant internal methods (\code{get}, \code{set}, \code{has}, \code{apply} and \code{construct}).
Existing native objects (such as \code{JSON}) are wrapped in a mock when first requested by the program, and have their calls and properties proxied to the original object.
Undefined properties are filled with \textit{full mocks}.
Full mocks are objects that try to be as generic as possible, since we do not know the proper value that the program expects them to have.
These mocks $(i)$ have an infinite number of properties, $(ii)$ are instantiable, $(iii)$ are iterable, $(iv)$ are invocable, and $(v)$ are thenable.
Another feature of mocks is that they will try to invoke any callback they receive in the arguments of a function call.
For example, the one passed to \code{window.addEventListener("load", cb)}.
This is done to cover as many code paths as possible, given that the main design goal of Fakeium is to complement static analysis.
As such, we are interested in exploring all possible behaviors of a program.

\vspace{1mm}
\noindent\textbf{Hooks.}\xspace
In some cases, mocking all encountered variables inside the sandbox may not be good enough to accurately simulate the runtime environment a program expects.
Examples of these cases include programs that expect a fixed value for a given variable, or that require an external payload to be fetched from the network.
For these cases, Fakeium supports great extensibility of its sandbox, allowing developers to hook custom variables into it.
Hooks are designed with usability and versatility in mind, supporting three different types of payloads: serializable values, functions and references.
Listing~\ref{lst:hook-examples} shows code snippets for all of them.

\begin{lstlisting}[
  float=t,
  caption={Examples of hooks supported by Fakeium.},
  label={lst:hook-examples},
  language=JavaScript
]
import { Reference } from "fakeium";

// Serializable value hook
fakeium.hook("document", {
  nodeType: 9, // Node.DOCUMENT_NODE
  readyState: "complete"
});
fakeium.hook("require", undefined);

// Function hook
fakeium.hook("bridge", () => {
  console.log("Hi from outside the sandbox!");
  return { success: true }; // Copied to the sandbox
});

// Reference hook
fakeium.hook("chrome", new Reference("browser"));
\end{lstlisting}

Hooks that hold serializable values have their contents copied to the sandbox using the structured clone algorithm~\cite{mdn-structured-clone}.
This method is natively performed by the V8 engine and supports the most common JavaScript types, with the exception of functions.
Function hooks are bindings exposed inside the sandbox that call functions outside it.
These functions run in the main Node.js isolate and can take both arguments and return values (even promises), as long as they are serializable.
References are hooks that redirect \code{get}, \code{set} and \code{apply} traps to other variables \textit{inside} the sandbox.
Most notably, Fakeium uses them to rewire aliases of the \code{globalThis} object by default, such as \code{window}.

\vspace{1mm}
\noindent\textbf{Module Resolution.}\xspace
Besides running scripts independently, Fakeium also supports sources that import other modules through a \textit{module resolver}.
The resolver is a custom callback defined by the developer that takes a \code{URL} instance~\cite{mdn-url} and returns a \code{Buffer} with its respective source code, or \code{null} if it does not exist.
Prior to calling the resolver, Fakeium takes care of resolving relative specifiers found in \code{import} statements, so that it always receives an absolute URL argument.
By default, Fakeium uses the \textit{``file:///''} origin, but it supports any other compliant value such as \textit{``https://example.com''} or \textit{``chrome-extension://''}.

\vspace{1mm}
\noindent\textbf{Events.}\xspace
Fakeium keeps track of events that occur during the execution of a program, namely variable reads, variable writes and function calls.
Events are captured by the \code{Proxy} traps defined in mocks and sent from the sandbox to the main isolate as soon as they are emitted, rather than waiting until the end of the execution.
This approach ensures that Fakeium captures events even if the program crashes or throws an uncaught error.
For each event, the sandbox logs the location in the code where it originated (filename, line and column), which is especially useful for programs that spawn across several modules.
We use the internal V8 function \code{Error.captureStackTrace} to get this information without having to instantiate, throw and catch an error.

\vspace{1mm}
\noindent\textbf{Object Tainting.}\xspace
Fakeium makes extensive use of tainting to keep track of various aspects of a source's execution.
To taint objects, we assign them custom properties where the key is a \code{Symbol} instead of a string to reduce the memory footprint.
We hide these symbol properties from the running program at the \code{Proxy} traps to avoid altering its behavior.
We use tainting to
$(i)$ mark mock objects and avoid mocking them twice,
$(ii)$ assign unique identifiers to objects being logged in events,
$(iii)$ mark visited callbacks to prevent endless loops,
and $(iv)$ mark stub functions that originate from a function hook.

\subsection{Limitations and Future Work}

While we strive to make Fakeium as feature-rich as possible, some advanced features have been left out of the first stable release to prioritize performance and stability.
Most importantly, Fakeium does not currently support dynamic imports through \code{await import()} or \code{require()} due to a limitation of isolated-vm.
We note that this is not a permanent limitation as there is already a patch from the community that could make implementing dynamic imports feasible in a future update.\footnote{
  See \url{https://github.com/automata-mc/isolated-vm/commit/9404fcc}.
}
Furthermore, this limitation can currently be overcome by using hooks, as demonstrated in Section~\ref{sec:case-study}.

A second aspect that could be improved is our DOM emulation.
While Fakeium has safeguards to mitigate infinite loops that may occur during traversal (\eg by constantly getting a truthy \code{firstElementChild} value), a more advanced implementation might be better suited for certain cases.
We are aware of this and are committed to refining this functionality in future updates.

\section{Evaluation}
\label{sec:evaluation}

We evaluate Fakeium on various datasets of JavaScript programs and compare it to a state-of-the-art static analysis pipeline.
Although our tool intends to supersede dynamic analysis when constraints deem it necessary, we evaluate against static analysis instead, since
$(i)$ Fakeium has more in common with the latter (it tries to find all possible behaviors of a program),
and $(ii)$ we want to measure the time footprint of Fakeium relative to static analysis to see if it has a significant impact.
In addition, more than half of the related work tools are no longer maintained, require compiling long-outdated versions of Chromium, and only perform a very specific task (such as removing unused dependencies) instead of providing a full-fledged dynamic execution environment.
Overall, Fakeium has a small impact on analysis time and resources, and provides additional signals otherwise undetected by static analysis.

\subsection{Datasets}
We use three different datasets of JavaScript sources that are representative of the current landscape.

\vspace{1mm}
\noindent\textbf{Browser Extensions.}\xspace
We collected the CRX packages for the top \datasetExtensionsTotal extensions with the most user installs from the Chrome Web Store as of July 2024.
CRX packages are a variant of ZIP archives that contain all the source code and assets needed to run an extension, making it easier to distribute~\cite{crx3-design-doc}.
The Chrome Web Store (CWS) is Google's official distribution platform for extensions that run on Google Chrome~\cite{chrome-web-store}.
We note that other Chromium-based browsers like Microsoft Edge, Brave, and Opera can also install extensions from this source since they run on the same engine.

\vspace{1mm}
\noindent\textbf{JavaScript Libraries.}\xspace
Also in late July 2024, we crawled the latest version of all \datasetCdnjsTotal libraries served by cdnjs at that time.
cdnjs is the most popular Content Delivery Network (CDN) for JavaScript assets, powered by Cloudflare and operating as an open-source project~\cite{cdnjs}.

\vspace{1mm}
\noindent\textbf{Obfuscated Scripts.}\xspace
To evaluate Fakeium's resistance to obfuscation, we wrote a simple script with 10 function calls executed in different ways, and generated obfuscated versions of it.
We used javascript-obfuscator~\cite{javascript-obfuscator}, the most popular free and open source tool of its kind, and produced a source file for each of its presets (default, low, medium, and high obfuscation).
We also added a stronger custom configuration based on the high preset, where we lowered the \code{splitStringsChunkLength} to 3 characters, increased the \code{stringArrayWrappersCount} to 8 (above the recommended limit of 5), and enabled unicode escape sequences.
Listing~\ref{lst:obfuscation-base} (Appendix) shows the source code for the base script used to generate the obfuscated versions.

\subsection{Methodology}

\begin{figure}[t]
  \centering
  \includegraphics[width=\columnwidth]{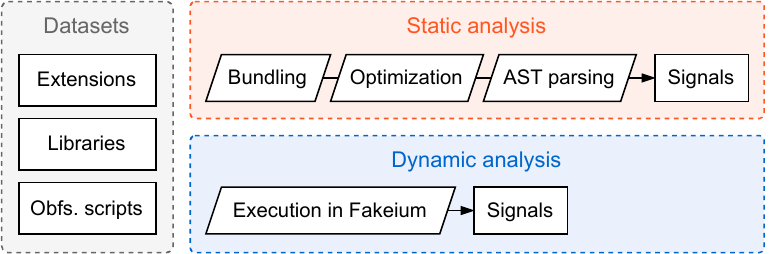}
  \caption{Evaluation methodology pipeline.}
  \label{fig:methodology}
\end{figure}

We follow the methodology pipeline shown in Figure~\ref{fig:methodology}.
We analyze the resources in each dataset both statically and dynamically.
We run the analyses sequentially (without parallelizing) on a modest server with two Intel Xeon E5645 CPUs and 16 GB of RAM.

For browser extensions, we identify and analyze all entrypoints found in the manifest file~\cite{chrome-manifest} and HTML documents that contain JavaScript code.
As mentioned in Section~\ref{sec:background:static-analysis}, a significant limitation of static analysis is that it cannot perform module resolution, which is used by several browser extensions in our dataset.
To overcome this, we use esbuild~\cite{esbuild} to bundle the various files imported from an entrypoint into a single JavaScript source.
We also use this library to apply some optimizations to the input files that will ease the forthcoming AST parsing, specifically tree shaking and identifier and syntax minification.
Lastly, we generate and traverse the AST using Babel~\cite{babeljs} to extract API calls and all string literals.

For dynamic analysis, we solely use Fakeium to run the dataset sources, and configure it with a timeout of 10 seconds and a heap memory limit of 256 MiB.

We do not perform any bundling or optimization before running them because the sandbox already supports module resolution (see Section~\ref{sec:fakeium:implementation}).
We extract API calls and string literals found in function arguments from the report that Fakeium generates.

To evaluate performance, we measure the wall time and heap memory usage of the AST generation and traversal phases and compare them to metrics from the sandbox execution.
For the static analysis, we additionally measure the wall time for the bundling and optimization phases.
Also in the case of static analysis, we calculate the memory usage as the difference between the used heap size before and after invoking the AST generation program, forcibly calling the V8 garbage collector to get more reliable results.
In the case of dynamic analysis, Fakeium has built-in support for tracking time and memory within the sandboxed V8 isolate, which we measured just before the end of the source execution.

To evaluate coverage, we introduce a new metric called the number of \textit{hidden API calls}.
These are API calls which are found by Fakeium but are missing from the static analysis.
We evaluate a similar metric over strings, and use a combination of regular expressions and later manual analysis to group them by category (\eg URLs, Google Analytics IDs).

We discarded extensions that had no entrypoints and any sources that caused esbuild to throw an error during static analysis, ending up with \datasetExtensionsOk extensions (\datasetExtensionEntrypointsOk entrypoints) and \datasetCdnjsOk cdnjs libraries.
As for the dynamic analysis, we note that Fakeium crashed due to an uncaught error for \datasetExtensionEntrypointsDynamicCrashP of extension entrypoints and \datasetCdnjsDynamicCrashP of libraries.
It also timed out in \datasetExtensionEntrypointsDynamicTimeoutP of entrypoints and \datasetCdnjsDynamicTimeoutP of libraries.
While the number of crashed sources is somewhat high, this is expected due to the mocking nature of Fakeium, as some intentionally throw an error if certain environmental conditions are not met.
Nevertheless, we do not discard these sources because Fakeium generates a report even when the sandbox crashes or times out.

\subsection{Performance}
\begin{figure}[t]
  \centering
  \includegraphics[width=\columnwidth]{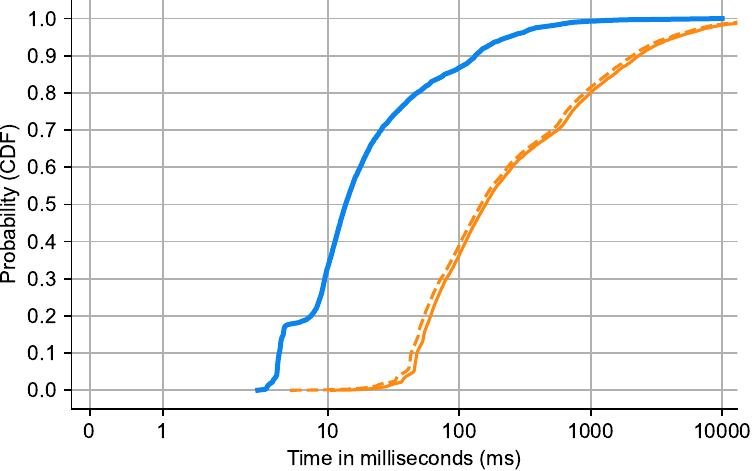}
  \caption{%
    Wall time distributions for static and dynamic analysis pipelines.
    \textcolor{orange}{Solid orange} line for static analysis (whole duration),
    \textcolor{orange}{dashed orange} line for static analysis (only AST parsing),
    \textcolor{blue}{\textbf{bold solid blue}} line for Fakeium.
  }
  \label{fig:performance-time}
\end{figure}
Figure~\ref{fig:performance-time} shows three different curves comparing the wall time of the static analysis pipeline against Fakeium.
For the static analysis, we plot one curve covering the entire process (bundling, optimization and AST parsing) and another for the AST parsing phase time only.
As can be appreciated, there is not much difference between these two due to esbuild being generally fast.
On average, our static analysis pipeline takes \perfStaticTimeAvg ms to run, with a standard deviation of \perfStaticTimeStd ms.
We see a substantial difference when running the sources from our datasets in Fakeium.
In this case, execution times are orders of magnitude lower, with an average of \perfDynamicTimeAvg ms and a standard deviation of \perfDynamicTimeStd ms.
Although we configure the sandbox with a hard timeout of 10 seconds for operational reasons, it had almost no effect on the results since the 99\% quantity is found at the \perfDynamicTimeQNN ms mark.
Overall, we can confidently say that Fakeium has a minimal time overhead when run alongside static analysis.

\begin{figure}[t]
  \centering
  \includegraphics[width=\columnwidth]{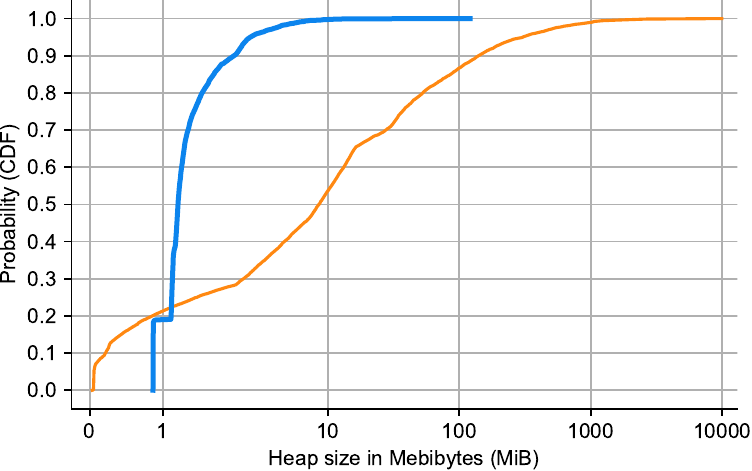}
  \caption{%
    Heap memory size distributions for static and dynamic analysis pipelines.
    \textcolor{orange}{Orange} line for static analysis,
    \textcolor{blue}{\textbf{bold blue}} line for Fakeium.
  }
  \label{fig:performance-memory}
\end{figure}
Figure~\ref{fig:performance-memory} compares the heap memory size of AST generation and traversal (parsing) in the static analysis pipeline with that of the V8 isolate sandbox spawned by Fakeium.
ASTs generated by Babel take up a decent amount of memory, reaching \perfStaticMemoryMax MiB for one of the sources in our dataset.
However, they can be extremely optimized for small programs, taking just a few bytes in such instances.
As such, the average heap size for ASTs is \perfStaticMemoryAvg MiB with a wider standard deviation of \perfStaticMemoryStd MiB.
An interesting observation that can be gleaned from this figure is that Fakeium always requires at least \perfDynamicMemoryMin MiB of heap.
This space includes the minimal V8 isolate footprint and the allocated size for holding the code and data structures of the bootstrap script (see Section~\ref{sec:fakeium:architecture}).
Regardless, Fakeium still uses much less memory than static analysis, with an average heap size of \perfDynamicMemoryAvg MiB and a standard deviation of \perfDynamicMemoryStd MiB.
The worst case we found was a source from our dataset that allocated \perfDynamicMemoryMax MiB of heap space (still well below the top bound we encountered in static analysis).
These numbers make our tool suitable for environments with low memory limitations.

\subsection{Signals}

\vspace{1mm}
\noindent\textbf{API Calls.}\xspace
We measure the number of API calls found by each analyzer and identify hidden API calls, \ie API calls observed by Fakeium that are not reported through static analysis.
These hidden API calls are particularly interesting as they demonstrate Fakeium's ability to uncover behaviors that would otherwise be missed.
On average, the static analyzer finds \apiCallsStaticAvgStd API calls per source, while Fakeium reports a lower number of \apiCallsDynamicAvgStd items.
We believe this is caused by unreachable code that never gets executed in the sandbox, and to a lesser extent by pruning API calls ending in \code{toString} and \code{valueOf}, which we do to ignore trivial and internal V8 calls.
To us, this is an expected result as static analysis tends to extract more signals than dynamic, except for obfuscated sources.

\begin{figure}[t]
  \centering
  \includegraphics[width=\columnwidth]{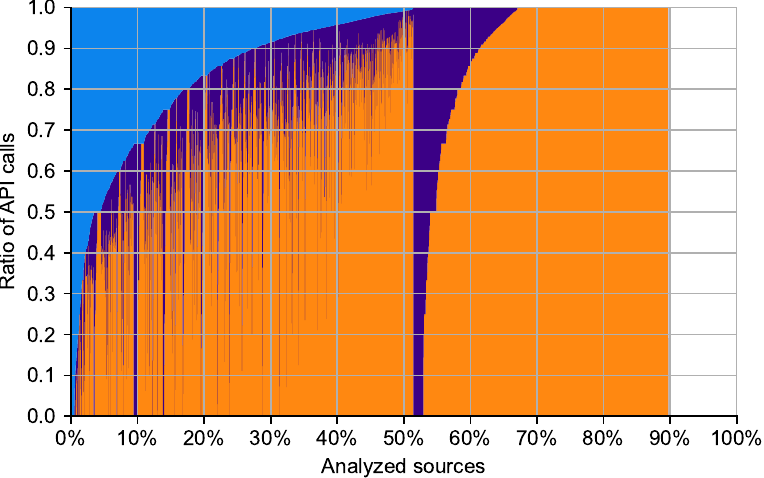}
  \caption{%
    Ratio of API calls per source, grouped by the analyzer that found them.
    \textcolor{orange}{Orange} area for static analysis only,
    \textcolor{blue}{blue} for Fakeium only (hidden API calls),
    \textcolor{purple}{purple} for found by both.
    Empty area for sources with no API calls.
  }
  \label{fig:api-calls}
\end{figure}
Figure~\ref{fig:api-calls} shows the ratio of API calls found per source, grouped by whether they are found only by the static analyzer, only by Fakeium (\ie hidden API calls), or by both.
Each vertical bar of pixels represents an analyzed source, sorted from left to right by the largest hidden API call ratio for better visualization.
We find \apiCallsEmptyExtensions extensions and \apiCallsEmptyLibraries cdnjs libraries that have no API calls for either static or dynamic analysis.
We attribute this to extensions with HTML entrypoints that have no source code, and small library plugins that do not access variables from the global scope (thus not registering as proper API calls).
An example of the latter is ember-computed-reverse,\footnote{
  See \url{https://cdnjs.com/libraries/ember-computed-reverse}.
} a tiny utility that provides a macro for the Ember.js framework~\cite{emberjs} to reverse arrays on build.
This leaves us with a total of \apiCallsNotEmptySources sources with detected API calls, with \apiCallsSourcesWithHiddenP of them having at least one hidden API call that was found by Fakeium but missing from static analysis.
On average, our tool detects \apiCallsHiddenAvgStd hidden API calls per analyzed source.

We also take a qualitative look at API calls that are relevant from a security and privacy perspective, which are listed and grouped by category in Table~\ref{tab:api-calls}.
Fakeium is specially good at detecting network-related invocations that are missed during static analysis, both for the legacy \code{XMLHttpRequest} and modern \code{fetch} objects.
Similarly, it logs the registration of event listeners commonly used in browser extensions to monitor the websites visited by the user, such as \code{browser.tabs.onUpdated}.

\begin{table}[t]
  \centering
  \caption{Sample of security and privacy API calls found by Fakeium.}
  \label{tab:api-calls}
  \resizebox{\columnwidth}{!}{%
    \begin{tabular}{lrrr}
      \toprule
      \thead[l]{API Call} & \thead[r]{Static} & \thead[r]{Hidden} & \thead[r]{Diff.} \\
      \midrule
      \multicolumn{4}{l}{\cellcolor[HTML]{EFEFEF}Code Execution} \\
      browser.runtime.setUninstallURL & 287 & 18 & +6\% \\
      browser.scripting.registerContentScripts & 56 & 2 & +4\% \\
      browser.scripting.executeScript & 422 & 1 & -- \\
      browser.tabs.executeScript & 144 & 0 & -- \\
      \multicolumn{4}{l}{\cellcolor[HTML]{EFEFEF}Network Traffic} \\
      XMLHttpRequest & 1.7k & 546 & +33\% \\
      browser.webNavigation.onBeforeNavigate.ad... & 36 & 5 & +14\% \\
      browser.webRequest.onBeforeRequest.addLis... & 116 & 10 & +9\% \\
      fetch & 1.9k & 12 & +1\% \\
      \multicolumn{4}{l}{\cellcolor[HTML]{EFEFEF}Privacy} \\
      browser.downloads.onCreated.addListener & 7 & 3 & +43\% \\
      browser.omnibox.onInputChanged.addListene... & 6 & 1 & +17\% \\
      browser.downloads.onChanged.addListener & 23 & 2 & +9\% \\
      browser.tabs.onUpdated.addListener & 402 & 24 & +6\% \\
      browser.omnibox.onInputEntered.addListene... & 19 & 1 & +5\% \\
      browser.cookies.get & 89 & 2 & +2\% \\
      browser.tabs.query & 1.1k & 24 & +2\% \\
      browser.cookies.getAll & 113 & 1 & +1\% \\
      navigator.mediaDevices.getUserMedia & 201 & 0 & -- \\
      navigator.geolocation.getCurrentPosition & 25 & 0 & -- \\
      navigator.mediaDevices.enumerateDevices & 98 & 0 & -- \\
      browser.management.getAll & 52 & 0 & -- \\
    \bottomrule
    \end{tabular}
  }
\end{table}

Given its extremely low overhead, we find that Fakeium is worth running alongside static analysis to complement its signals, as it detects API calls that even complex static analysis setups with bundling support miss.

\vspace{1mm}
\noindent\textbf{Strings.}\xspace
As with API calls, the static analyzer finds more strings on average than the dynamic analyzer (\stringsStaticAvg versus \stringsDynamicAvg).
Note, however, that this is expected since the static analyzer logs all strings found in the AST, while Fakeium only keeps track of those that are involved in an API call (\eg passed as arguments).
Despite this significant volume difference, Fakeium still successfully uncovers hidden strings in \stringsSourcesWithHiddenP of the dataset sources.

Looking at the \stringHiddenTotal unique hidden strings uncovered by our tool, we find some interesting entries.
These include \stringHiddenIpvFour hardcoded IPv4 ranges used by VPN extensions, \stringHiddenUuid different UUIDs, \stringHiddenUrls URLs, and \stringHiddenJson JSON documents.
Most of the aforementioned JSON documents are network request payloads, with at least \stringHiddenJsonUuid of them containing unique identifiers and timestamps that are likely used for tracking analytical events.
Among the URLs missed by the static analyzer, we find \stringHiddenUrlsOrigins different origins and \stringHiddenUrlsGa Google Analytics measurement IDs~\cite{google-analytics-measurement-id}.
These hidden strings bring meaningful insight to security analysts by providing unique identifiers that can be used to characterize malware.

\subsection{Analyzing Obfuscated Sources}
Table~\ref{tab:obfuscation} lists the detected API calls and string literals found by the static analyzer and our tool for the different versions of the obfuscated sources.
Both analyzers perfectly spot all 10 calls and 10 strings for the low-level obfuscation configurations.
However, static analysis has difficulties with the medium and high presets, whereas our tool manages to maintain the same accuracy at all levels.

Given the results our experiments, we find Fakeium suitable for analyzing obfuscated sources.

\begin{table}[t]
  \centering
  \caption{Comparison of the number of API calls and strings detected~in~obfuscated~sources.}
  \label{tab:obfuscation}
    \begin{tabular}{lrrrr}
      \toprule
      \multirowthead{2}[-2pt]{Level} & \multicolumn{2}{l}{\thead[c]{Static}} & \multicolumn{2}{l}{\thead[c]{Fakeium}} \\
                                     & \thead[l]{Calls} & \thead[l]{Strings} & \thead[l]{Calls} & \thead[l]{Strings}  \\
      \midrule
      None & 10\textcolor{gray}{\scriptsize/10} & 10\textcolor{gray}{\scriptsize/10} & 10\textcolor{gray}{\scriptsize/10} & 10\textcolor{gray}{\scriptsize/10} \\
      Default & 10\textcolor{gray}{\scriptsize/10} & 10\textcolor{gray}{\scriptsize/10} & 10\textcolor{gray}{\scriptsize/10} & 10\textcolor{gray}{\scriptsize/10} \\
      Low & 10\textcolor{gray}{\scriptsize/10} & 10\textcolor{gray}{\scriptsize/10} & 10\textcolor{gray}{\scriptsize/10} & 10\textcolor{gray}{\scriptsize/10} \\
      Medium & 5\textcolor{gray}{\scriptsize/10} & 2\textcolor{gray}{\scriptsize/10} & 10\textcolor{gray}{\scriptsize/10} & 10\textcolor{gray}{\scriptsize/10} \\
      High & 2\textcolor{gray}{\scriptsize/10} & 0\textcolor{gray}{\scriptsize/10} & 10\textcolor{gray}{\scriptsize/10} & 10\textcolor{gray}{\scriptsize/10} \\
      Custom & 2\textcolor{gray}{\scriptsize/10} & 0\textcolor{gray}{\scriptsize/10} & 10\textcolor{gray}{\scriptsize/10} & 10\textcolor{gray}{\scriptsize/10} \\
    \bottomrule
    \end{tabular}
\end{table}

\section{Case Study}
\label{sec:case-study}

In this section, we showcase with a practical example how Fakeium can be applied to the analysis of a recent in-the-wild malware sample.
Back in August 2024, researchers at Checkmarx uncovered a campaign targeting npm with dozens of malicious packages that mimicked ``noblox.js,'' a popular library for interacting with the Roblox API~\cite{report-npm-noblox}.
The malicious payload hides in the ``postinstall.js'' file of the impersonating packages, of which we were able to obtain an archived sample from Socket.\footnote{
  Available at \url{https://socket.dev/npm/package/noblox.js-async/files/4.6.9/postinstall.js}.
}
This script is heavily obfuscated and crashes when trying to call \code{require()}, which is \code{undefined} by default in Fakeium.
Statically analyzing this source does not provide any meaningful signals about its behavior.

To overcome this issue, we modify the value of \code{globalThis.require} using our tool's hooking capabilities.
This allows us to mock all the modules required by the malicious script, and even log and modify arguments and return values.
In particular, we safely create \textit{ad hoc} mocks that simulate the ``child\_process'', ``crypto'', ``fs'', ``node-fetch'', ``os'' and ``path'' modules to prevent the program from crashing.
We also return fixed values for \code{os.userInfo().username} and \code{crypto.randomBytes()} to taint their outputs, and implement the minimum functionality of the missing \code{Buffer} class.
Listing~\ref{lst:hook-example} shows a simplified version of the hooks we use.

\begin{lstlisting}[
  float=h,
  caption={Hooks used to analyze the case study malware (simplified).},
  label={lst:hook-example},
  language=JavaScript
]
fakeium.hook("require", specifier => {
  if (specifier === "crypto") {
    return { randomBytes: () => "fixedValue" };
  }
  if (specifier === "os") {
    return {
      userInfo: () => {
        return { username: "testUsername" };
      }
    };
  }
  if (specifier === "child_process") {
    return {
      execSync: cmd => {
        console.log("execSync()", cmd);
      }
    };
  }
  // [...]
});

fakeium.hook("Buffer.from", (value, encoding) => {
  return {
    toString: () => Buffer.from(value, encoding).toString()
  };
});
\end{lstlisting}

With just these minor customizations to the Fakeium sandbox, we successfully run the sample and infer its behavior.
The analysis report shows the first stage of ``postinstall.js'' performing multiple calls to \code{parseInt} and \code{setInterval} to unpack the malicious payload.
Then, it uses ``node-fetch'' to download an executable from a GitHub repository\footnote{
  Hosted at \url{https://github.com/aspdasdksa2/callback/raw/main/cmd.exe}.
} and saves it to disk with a random filename.
Afterwards, it adds several keys to the Windows Registry using \code{execSync()} to bypass UAC and execute the downloaded binary with elevated privileges (Listing~\ref{lst:uac-bypass}).
Finally, it sends a message to a Discord channel with the body ``someone got trolled, Uac Bypassed, Quasar Ran'', presumably notifying the malware operators of the successful execution.

\begin{lstlisting}[
  float=h,
  caption={UAC bypass via FodHelper.exe commands (simplified).},
  label={lst:uac-bypass},
  language=Bash
]
reg add "HKCU\Software\Classes\AppX82a6gwre4fdg3bt635tn5ctqjf8msdd2\Shell\open\command" /f
reg add "HKCU\Software\Classes\AppX82a6gwre4fdg3bt635tn5ctqjf8msdd2\Shell\open\command" /ve /t REG_SZ /d "C:\WindowsApi\fixedValue.exe" /f
reg add "HKCU\Software\Classes\ms-settings\Shell\Open\command" /f
# [...]
reg add "HKCU\Software\Classes\ms-settings\Shell\Open\command" /v DelegateExecute /t REG_SZ /d "" /f
start "" "C:\Windows\System32\fodhelper.exe"
\end{lstlisting}

This case study demonstrates Fakeium's extensive customization options, ease of use, and overall ability to analyze recent malware, even when heavily obfuscated.

\section{Conclusion}
\label{sec:conclusion}

Dynamic analysis of Javascript applications (such as browser extensions) is a challenging problem that consumes significant time and computing resources, making it difficult to scale.
In this paper, we presented Fakeium, a lightweight sandbox based on the V8 engine for executing untrusted JavaScript code.
Fakeium is designed as a substitute to traditional dynamic analysis for when the latter is not feasible.
Through extensive experimentation, we demonstrate that Fakeium exhibits a tiny overhead of just \perfDynamicTimeAvg ms to run an average program in \perfDynamicMemoryAvg MiB of heap, making it an excellent fit to run alongside static analysis due to its minimal impact on the existing pipeline.
It is also highly customizable, allowing developers to create custom hooks inside the sandbox for more complex instrumentations.
Furthermore, the dynamic nature of Fakeium makes it quite resistant to obfuscation, proving capable of extracting signals from sources that have been subject to several transformations.

\bibliographystyle{IEEEtran}
\bibliography{paper}

\appendix

Listing~\ref{lst:obfuscation-base} provides the source code of the script used as a base to generate the obfuscated sources dataset.

\begin{lstlisting}[
  float=ht,
  caption={Base script used for obfuscation experiments.},
  label={lst:obfuscation-base},
  language=JavaScript
]
// Regular function calls
firstFn("ARG_1");
secondFn("ARG_2");

// Async function calls
(async () => {
  thirdFn("ARG_3");
  await fourthFn("ARG_4");
  fifthFn("ARG_5");
})();

// Callbacks
window.addEventListener("load", () => {
  sixthFn("ARG_6");
});
setTimeout(() => {
  seventhFn("ARG_7");
  eighthFn("ARG_8");
}, 1000);

// More complex logic
(() => {
  for (let i=0; i<Number.MAX_VALUE; i++) {
    if (i === 20) {
      ninthFn("ARG_9");
      break;
    }
  }
  tenthFn("ARG_10");
})();
\end{lstlisting}

\vfill
\end{document}